\documentclass[useAMS,a4paper,usenatbib]{mn2e}

\usepackage{graphicx, array, amssymb,amsmath}

\usepackage{rotating}
\usepackage{lscape}
\usepackage{verbatim}

\usepackage{tabularx,colortbl}
\newcommand{\be}{\begin{equation}}
\newcommand{\ee}{\end{equation}}
\newcommand{\bea}{\begin{eqnarray}}
\newcommand{\eea}{\end{eqnarray}}

\newcommand\lsim{\mathrel{\rlap{\lower4pt\hbox{\hskip1pt$\sim$}}
    \raise1pt\hbox{$<$}}}
\newcommand\gsim{\mathrel{\rlap{\lower4pt\hbox{\hskip1pt$\sim$}}
    \raise1pt\hbox{$>$}}}
\def\ee{\end{equation}}
\def\be{\begin{equation}}
\newcommand{\Omk}{\Omega_\kappa}

\newcommand{\mc}[1]{{\mathcal{#1}}}
\newcommand{\data}{d}
\newcommand{\params}{\theta}

\newcommand{\pzero}{\psi}
\newcommand{\mdl}{\mc{M}}

\newcommand{\dr}{\rm{d}}

\newcommand{\Ode}{\Omega_\text{de}}

\newcommand{\zbao}{z_\text{bao}}

\newcommand{\curv}{R_c}

\begin{document}
\setlength{\unitlength}{1mm}

\title[Applications of Bayesian model averaging]{Applications of Bayesian model averaging to the curvature and size of the Universe}

\author[Vardanyan, Trotta and Silk]{Mihran Vardanyan$^{1}$\thanks{E-mail:
mva@astro.ox.ac.uk}, Roberto Trotta$^{2}$\thanks{E-mail:
r.trotta@imperial.ac.uk} and Joseph Silk$^{1}$\thanks{E-mail:
silk@astro.ox.ac.uk}\\
$^{1}$Oxford University, Astrophysics,  Denys Wilkinson Building,
Keble Road, Oxford, OX1 3RH, UK \\
$^{2}$Imperial College London, Astrophysics Group, 
	  Blackett Laboratory, Prince Consort Road, London SW7 2AZ, UK}

%\date{Accepted 1988 December 15. Received 1988 December 14; in original form 1988 October 11}

%\pagerange{\pageref{firstpage}--\pageref{lastpage}} \pubyear{2002}

\maketitle

\begin{abstract}
Bayesian model averaging is a procedure to obtain parameter constraints that account for the uncertainty about the correct cosmological model. We use recent cosmological observations and Bayesian model averaging to derive tight limits on the curvature parameter, as well as robust lower bounds on the curvature radius of the Universe and its minimum size, while allowing for the possibility of an evolving dark energy component. Because flat models are favoured by Bayesian model selection, we find that model-averaged constraints on the curvature and size of the Universe can be considerably stronger than non model-averaged ones. For the most conservative prior choice (based on inflationary considerations), our procedure improves on non model-averaged constraints on the curvature by a factor of $\sim 2$. The curvature scale of the Universe is conservatively constrained to be $\curv > 42$ Gpc ($99\%$), corresponding to a lower limit to the number of Hubble spheres in the Universe $N_U > 251$ ($99\%$).
\end{abstract}

\begin{keywords}
cosmology: theory; cosmology:
cosmological parameters; methods: statistical.
\end{keywords}

\section{Introduction}

One of the most spectacular advancements of observational cosmology over the past two decades has been the ability to measure the spatial geometry of the universe with unprecedented accuracy.  In the Friedmann-Robertson-Walker (FRW) Universe there are only three discrete possibilities for the underlying geometry, namely flat, open or closed. The amount of curvature is usually characterized by the curvature parameter $\Omk$: if $\Omk < 0$ the geometry of spatial sections is spherical (i.e., the Universe is closed) and the Universe has a finite size. If instead $\Omk>0$ the geometry is hyperbolic (i.e., the Universe is open), while for $\Omk=0$ spatial sections are flat. In both the two latter cases, the spatial extent of the Universe is infinite. 

Limits on the value of $\Omk$ can be derived in a geometrical way by observing the angular size subtended by cosmological features of known physical length, such as the acoustic peaks in the cosmic microwave background (CMB) and the corresponding baryonic acoustic oscillations (BAO) in the distribution of large scale structures. Furthermore, type Ia supernovae (SNIa) can be used as standard candles to determine the luminosity distance as a function of redshift. A combination of these three probes has been succesfully used to set very tight limits to the curvature parameter, which is now constrained at better than the $\sim 10^{-3}$ level. For example, ~\cite{Komatsu:2010fb} find
 $ \Omk = -0.0057^{+0.0066}_{-0.0068} $ at 68 \% CL, employing a combination of WMAP7, BAO~\citep{Percival:2009xn}  and SNIa data~\citep{Hicken:2009dk}. 
 %)~\citep{Komatsu:2010fb} and $ \Omk = -0.0050^{+0.0061}_{-0.0060} $ (68 \% CL from WMAP5+BAO~\citep{Percival:2007yw} +SN (UNION08~\cite{Kowalski:2008ez}) ~\citep{Komatsu:2008hk}. 

Impressive as such limits are, they {\em assume} the Universe to be curved, and carry out parameter inference on the quantity describing curvature. A different methodological perspective is required to go beyond that assumption: Bayesian model comparison can be used to obtain the posterior probability of the Universe being curved~\citep{Vardanyan:2009ft}. The purpose of this paper is to further expand the approach adopted in~\cite{Vardanyan:2009ft} by deriving model-averaged limits on the curvature of the Universe, which fully account for the uncertainty in selecting the correct model for the FRW Universe. Given current data, flat models are preferred by Bayesian model selection from an Occam's razor perspective, and therefore most of the probability mass becomes concentrated in models with vanishing spatial curvature. As we demonstrate, model-averaged limits on the curvature parameter are {\em tighter} than non model-averaged constraints, because of this ``concentration of probability'' effect. 

This paper is organized as follows: in section~\ref{sec:setup} we outline the Bayesian model averaging framework. We describe the data sets used for the analysis and the models included in section~\ref{sec:models}. We then present our results in section~\ref{sec:results}, where we also discuss the dependency on prior choice. We give our conclusions in section~\ref{sec:conclusions}.

%%%%%%%%%%%%%%%%%%%%%%%%%%%%%%%%%%%%%%%%%%%%%%%%
\section{Bayesian model averaging}
\label{sec:setup}
%%%%%%%%%%%%%%%%%%%%%%%%%%%%%%%%%%%%%%%%%%%%%%%%

%Bayesian inference relies on Bayes' theorem, which expresses the posterior probability for a hypothesis $H$ after one has seen the data, $p(H|d)$,  to the prior probability for the hypothesis, $p(H)$ and the likelihood, $p(d|H)$:
%\be \label{eq:bayes}
%p(H|d) = \frac{p(d|H)p(H)}{p(d)}.
%\ee
If one is interested in determining the parameters $\theta$ of a given model $\mdl$, then the relevant quantity is the posterior distribution for $\theta$ under that model, which is given by Bayes theorem as 
\be \label{eq:level1}
p(\theta|d, \mdl) = \frac{p(d|\theta, \mdl)p(\theta | \mdl)}{p(d | \mdl)},
\ee
where the explicit conditioning on $\mdl$ indicates that the posterior pdf for $\theta$ given data $d$, $p(\theta|d, \mdl)$, is conditional on having assumed a specific model $\mdl$. This is the usual parameter inference step, and often the first level of inference in a problem (i.e., determining the constraints on the model's parameters).

The second level of inference is Bayesian model comparison, which aims to determine the relative probability of models themselves. The posterior probability of a model $\mc{M}_i$ given
the data, $p(\mc{M}_i|d)$ is related to the Bayesian evidence (or model
likelihood) $p(d|\mc{M}_i)$ by
\begin{equation} \label{eq:modelpost}
 p(\mc{M}_i|d) = \frac{p(d|\mc{M}_i)p(\mc{M}_i)}{p(d)}\, ,
\end{equation}
where $p(\mc{M}_i)$ is the prior for model $\mc{M}_i$, $p(d)=\sum_i
p(d|\mathcal{M}_i)p(\mc{M}_i)$ is a normalization constant (where the sum runs over all available models) and
\be \label{eq:Bayesian_evidence}
p(d|\mathcal{M}_i)=\int\!d\theta\, p(d|\theta, \mc{M}_i) p(\theta|\mc{M}_i)
\ee
is the Bayesian evidence, which appears as a normalization factor in Eq.~\eqref{eq:level1}. Given two competing models $\mc{M}_i,
\mc{M}_j$, the change in their relative probability in going from the prior (i.e, before we see the data) to the model posterior (after the data have been taken into account via the likelihood) is given by the Bayes factor $B_{ij}$:
\begin{equation}
 B_{ij} \equiv \frac{p(d|\mc{M}_i)}{p(d|\mc{M}_j)}\, ,
\end{equation}
where large (small) values of $B_{ij}$ denote a preference for $\mc{M}_i$ ($\mc{M}_j$). The `Jeffreys' scale' gives an empirical scale for
translating the values of $\ln B_{ij}$ into strengths of belief, with thresholds $|\ln B_{ij}| = 1.0, 2.5, 5.0$ separating levels of inconclusive, weak, moderate and strong evidence, respectively, see e.g.~\cite{Trotta:Bayes}. Recently, the framework of model comparison has been extended to include the possibility of `unknown models' discovery~\citep{doubt,March:2010ex}.

The third level of inference is represented by Bayesian model averaging (BMA), whose purpose is to determine constraints on common parameters among the models being considered ($\mdl_i$, with $i=1,\dots,N$) accounting for the uncertainty in selecting the correct model. This is the most general inference one can obtain on the parameters values (at least as long as the list of models is reasonably complete). The model-averaged posterior distribution for parameter $\theta$ is given by 
 \begin{align}
p(\theta \vert d) & = \sum_{i=1}^N p(\theta | d, \mdl_i) p(\mdl_i | d) \\ 
&= p(\mdl_1|d)  \sum_{i=1}^N B_{i1} p(\theta | d, \mdl_i)  \label{eq:BMA},
 \end{align}
where in the second equality we have replaced the models posterior probabilities, $p(\mdl_i | d)$ by the Bayes factors with respect to a reference model, here $\mdl_1$, and further assumed that the prior probabilities for the $N$ models are all equal, i.e.~$p(\mdl_i) = 1/N$, ($i=1,\dots, N$). With this assumption, the posterior for $\mdl_1$ is given by 
\begin{equation}
p(\mdl_1|d) = \frac{1}{1 + \sum_{i=2}^N B_{i1}}.
\end{equation}
BMA has been applied to the dark energy equation of state in~\cite{Liddle:2006kn}, to the scalar spectral index in~\cite{Parkinson:2010zr} and in the context of weak lensing and Sunyaev-Zel'dovich effect data in~\cite{Marshall:2003ez}.

\subsection{Computation of the Bayes factors}

Given two or more models, computing the Bayes factors entering Eq.~\eqref{eq:BMA} requires the evaluation of the multi-dimensional integral of Eq.~\eqref{eq:Bayesian_evidence}. Several algorithms are available today to compute the Bayesian evidence numerically. Here we are interested in the case where the models are {\em nested} within each other, i.e. where one of the models is obtained from a more complicated one for a speficic choice of some of the parameters of the latter. 
In our case, the extra parameters are the curvature, $\Omk$, and/or the dark energy equation of state parameters, $w_0$ or $w_a$, depending on the model under consideration. For example, a curved Universe reverts to a flat one for $\Omk=0$, or an evolving dark energy equation of state reverts to a cosmological constant model for $w_0 = -1, w_a = 0$. In this case, the Bayes factor between models $\mdl_ {i}$ and $\mdl_ {j}$ can be written in all generality as
 \begin{equation} \label{eq:savagedickey}
 B_{ij} = \left.\frac{p(\vartheta \vert \data, \mdl_{j})}{p(\vartheta |
 \mdl_  {j})}\right|_{\vartheta = \vartheta_*},
 \end{equation}
where we have split the more complicated model's parameters as $\params =
(\pzero, \vartheta)$, and $\vartheta$ are the extra parameters of model $\mdl_j$, which reduces to the simpler model $\mdl_i$ for $\vartheta = \vartheta_*$.
This expression is known as the Savage--Dickey density ratio
(SDDR, see \cite{Verdinelli:1995} and references therein. For cosmological applications, see~\cite{Trotta:2005ar}). The
numerator is simply the marginal posterior for $\vartheta$, evaluated at the value, $\vartheta = \vartheta_*$ (which can easily be obtained with standard Markov Chain Monte Carlo techniques), while the
denominator is the prior density for the extra parameters $\vartheta$ under the more complicated model,
evaluated at the same point. 

Once the Bayes factor for nested models which differ by one parameter at the time has been obtained using Eq.~\eqref{eq:savagedickey}, the Bayes factor between other models which have two or more nested parameters between them can be easily derived. If model $\mdl_i$ has one more parameter than model $\mdl_k$, which in turn has one more parameter than $\mdl_j$, the  Bayes factor between models $i$ and $j$ is given by 
 \begin{equation}
B_{ij}=B_{ik} \times B_{kj} ,
 \end{equation}
where the Bayes factor $B_{ik}$ and $B_{kj}$ can be obtained via the SDDR. A similar technique has been adopted in~\cite{Kunz:2006mc}. 
%%%%%%%%%%%%%%%%%%%%%%%%%%%%%%%%%%%%%%%%%%%%%%%%
\section{Models and data sets}
\label{sec:models}
%%%%%%%%%%%%%%%%%%%%%%%%%%%%%%%%%%%%%%%%%%%%%%%%

%We now present the models we consider in our model averaging procedure and the cosmological data sets adopted in our analysis. 

\subsection{Models and priors}

We work in the framework of FRW cosmologies including a cold dark matter (CDM) component, a possible curvature and a dark energy component. We will consider open ($\kappa = -1$) and closed ($\kappa = 1$) Universes as two separate models, as this allows us to adopt a prior on the curvature parameter $\Omk$ which is uniform in $\log(\Omk)$, as explained below. A model is fully specified by the choice of curvature and dark energy parameterization, alongside their respective priors. We consider the possibilities listed in Table~\ref{tab:models}: a model is defined by picking a choice from the upper part of the table (curvature) and one from the lower part (dark energy). From now on, a single model will be denoted by a pair of labels in the subscript, each referring to the prior choice for the curvature and dark energy sector. So, for example, the model $\mdl_{F\Lambda}$ denotes a flat Universe ($F$) with a cosmological constant ($\Lambda$). Correspondingly, Bayes factors between two models will have two pairs of labels in the subscript, separated by a comma for clarity, e.g.~$B_{OW,F\Lambda}$ denotes the Bayes factor between $\mdl_{OW}$ and $\mdl_{F\Lambda}$.

For non-flat models, we consider two different prior choices for the curvature parameter: a uniform prior in the range  $-1 \leq \Omk \leq 1$ (the `Astronomer's prior') and a uniform prior in the range $-5 \leq \log|\Omk| \leq 0$ (the 'Curvature scale prior'). The Astronomer's prior is motivated by basic consistency with observable properties of the Universe, such as the age of the oldest objects, while the Curvature scale prior is based on an inflationary scenario, see~\cite{Vardanyan:2009ft} for full details. For the dark energy equation of state, we adopt the parameterization 
\begin{equation} \label{eq:wz}
w(z) = w_0 + w_a z/(1+z),
\end{equation}
with two free parameters, $w_0, w_a$. Model $W$ has $w_a = 0$ and a uniform prior on $w_0$ in the range given in Table~\ref{tab:models}, while model $WZ$ further allows for $w_a \neq 0$. The prior ranges for $w_0, w_a$ are sufficiently wide to enclose the support of the likelihood function, but not too large in order to avoid a very strong Occam's razor effect against evolving dark energy models. Finally, parameters that are common to all models (such as the amplitude of primordial fluctuations or the baryonic density) are irrelevant for the model comparison, as shown by Eq.~\eqref{eq:savagedickey}, and therefore the choice of priors on them is unproblematic. For each choice of curvature prior, we thus consider a total of 9 cosmological models.
\begin{table*}
\begin{center}\begin{tabular}{| l | l l | }
\hline
Model & \multicolumn{2}{c|}{Parameters and priors} \\ 
\hline  & \multicolumn{2}{c|}{Curvature sector}  \\
& Astronomer's prior & Curvature scale prior
\\\hline
F (flat) & $\Omk = 0$ &  $\Omk = 10^{-5}$\\
O (open) & $0 \leq \Omk \leq 1$ (uniform)& $-5 \leq \log\Omk \leq 0$ (uniform)\\
C (closed) & $-1 \leq \Omk \leq 0$ (uniform)& $-5 \leq -\log\Omk \leq 0$ (uniform)\\
\hline  &  \multicolumn{2}{c|}{Dark energy sector}  \\\hline
$\Lambda$  &  \multicolumn{2}{l|}{$w_0 = -1, w_a = 0$ } \\
$W$ &  \multicolumn{2}{l|}{$-2 \leq w_0 \leq  -1/3$ (uniform), $w_a = 0$} \\
$WZ$&  \multicolumn{2}{l|}{$-2 \leq w_0 \leq  -1/3$ (uniform), $-1.33 \leq w_a \leq 1.33$ (uniform)} \\
\hline
\end{tabular} \caption{Prior choices for the curvature (top half) and dark energy (bottom half) parameters considered in the analysis. A model is fully specified by selecting a prior choice from the top and one from the bottom of the table, thus defining both the curvature and the dark energy sectors. \label{tab:models} }
\end{center}
\end{table*}

In the following, we will derive model-averaged constraints on the curvature parameter, $\Omk$, the curvature radius $\curv$, given by
\be 
 R_c   = \frac{c}{H_0}\frac{1}{|\Omk|^{1/2}} 
 \ee 
 (where $c$ is the speed of light and $H_0$ the Hubble constant today in km/s/Mpc) and the number of Hubble spheres in the Universe, $N_U$, defined as the ratio of the present volume of the spatial slice to the apparent particle horizon (see~\cite{Scott:2006kga} for details), 
\be
N_U \equiv \frac{2 \pi}{2 \chi - \sin(2\chi)},
\ee
where $\chi$ is the comoving radial distance. 

In some models, the value of some of the parameters is fixed. E.g., for flat models the curvature parameter vanishes, $\Omk=0$ (for the Astronomer's prior), while for open and flat model $\curv, N_U \rightarrow \infty$. In such cases, the posterior probability mass associated with that model in the model-averaged expression gets concetrated in a Dirac delta function $\delta$ around the fixed value of the parameter. So in the case of curvature, for example, the BMA expression~\eqref{eq:BMA} becomes:
\be
\begin{aligned}
p(\Omk \vert \data) = p(\mdl_{F\Lambda} | d) & {\Big [} \sum_{[i \neq F, j = \Lambda, W, WZ]} p(\Omk | d, M_{ij}) B_{ij, F\Lambda} +\\ 
& \sum_ {[j = \Lambda, W, WZ]} \delta(\Omk - \Omk^*) B_{Fj, F\Lambda} {\Big ]},
\end{aligned}
\ee
where the fixed value $\Omk^* = 0$ for the Astronomer's prior and $\Omk^* = 10^{-5}$ for the curvature scale prior. In the above expression we have taken the flat, $\Lambda$CDM model $F\Lambda$ as our reference model, and computed all Bayes factors with respect to it.
  
\subsection{Data sets employed}

The angular position of the first acoustic peak in the CMB power spectrum and the same acoustic signature in the correlation function of the galaxies provide us with standard rulers at high and low redshifts, respectively.   The measurements of the standard rulers in the direction perpendicular to the line of sight are used to constrain the angular diameter distance $D_A(z)=(1+z)^{-2}D_L(z)$, where $D_L$ is the luminosity distance, given by
\be \label{eq:lumdist}
D_{L}(z) = \frac{c}{H_0 \sqrt{|\Omk|}} (1+z) \sin \left( H_0 \sqrt{|\Omk|} \int_0^z \frac{\dr x}{H(x)}\right).
\ee
The Hubble function $H(z)$ is expressed in terms of the present-day matter-energy content of the Universe as follows: 
\begin{multline}
%\nonumber
H^2(z)= \Big( \Omega_m(1+z)^3+\Omega_r(1+z)^4+\Omk (1+z)^2\\
+\Ode \exp \left ( 3\int_{0}^{z}\frac{1+w(x)}{1+x} \dr x \right ) \Big).
\end{multline}
The dark energy time evolution is described by the present--day dark energy density in units of the critical density, $\Ode$, and by its equation of state, $w(z)$, as given in Eq.~\eqref{eq:wz}. In extracting constraints on cosmological parameters from luminosity or angular diameter distance measurements, one has to be careful to consider the potential impact of degeneracies between the assumed models. In this case, the strong degeneracy between curvature and dark energy evolution (see e.g.~\cite{Clarkson:2007bc}) is at least partially accounted for by admitting in our space of models an evolving dark energy equation of state. 

We include the WMAP 5--year data~\citep{Dunkley:2008ie} via their constraints on the shift parameters and the baryon density, following the method employed in~\cite{Komatsu:2008hk}. We notice that adopting WMAP 7--year data is not expected to change significantly our results, because of the fundamental geometrical degeneracy in the CMB. We  make use of the SDSS and 2dFGRS baryonic acoustic scale measurements following~\cite{Percival:2007yw}.  The scale of BAOs is used to estimate the quantity  
\be
D_V(z)= \left( c \zbao (1+z)^2 \frac{ D_A^2(\zbao)}{H(\zbao)}\right)^{1/3}, 
\ee
for  $\zbao=0.2$ and $\zbao=0.35$. We use two Gaussian data points with mean value and standard deviation $r_s/D_V(0.2)=0.1980 \pm 0.0058$ and $r_s/D_V(0.35)=0.1094\pm 0.0033$~\citep{Percival:2007yw}, where $r_s$ is the acoustic sound horizon. We also add the Hubble Key Project determination of the Hubble constant today, as a Gaussian datum with mean and standard deviation $H_0 = 72 \pm 8$ km/s/Mpc~\citep{Freedman:2000cf}. SN type Ia data are included in the form of the UNION08 data set sample~\citep{Kowalski:2008ez}. 

We employ a Metropolis--Hastings Markov Chain Monte Carlo procedure to derive the posterior distribution for the parameters in our model, and to compute the Bayes factors necessary for BMA via Eq.~\eqref{eq:savagedickey}. We take flat priors on the following quantities: $\Omega_m h^2, \Omega_b h^2, w_0, w_a$ (whenever $w_0$, $w_a$ is not fixed to $-1$, $0$, respectively). The prior bounds on the first 2 parameters are irrelevant, as the posterior is well constrained within the prior. The Bayes factors are obtained as the mean of the Bayes factors obtained from  of 8 independent reconstructions with MCMC, while their uncertainty is estimated from the variance of the values from the 8 runs.

\section{Results and Discussion}
\label{sec:results}

Table~\ref{tab:model_selection} gives the results of our model comparison between the flat $\Lambda$CDM model and the 8 alternative models considered, for both choices of curvature prior. As all values of $\ln B > 0$, the flat $\Lambda$CDM model is the one preferred by the data, as expected. The strength of evidence in its favour depends on the chosen prior, with the curvature scale prior giving in general a smaller evidence against the alternative models, since its Occam's razor effect on curvature is smaller than for the Astronomer's prior, as discussed in detail in~\cite{Vardanyan:2009ft}. For example, when comparing a flat $\Lambda$CDM model with a closed $\Lambda$CDM Universe, the odds in favour of the former are $\sim 100:1$ under the Astronomer's prior (moderate evidence), while only $2:1$ under the more conservative Curvature scale prior (inconclusive evidence). We also observe weak to moderate evidence against the evolving dark energy models as compared to models with a cosmological constant (both flat and curved). The posterior probability for flat models is given by 
\begin{equation}
p(\kappa = 0 | d) = \left(1 + \frac{\sum_{[i \neq F, j = \Lambda, W, WZ]} B_{ij, F\Lambda}}{\sum_ {[j = \Lambda, W, WZ]}B_{Fj, F\Lambda} }
\right)^{-1},
\end{equation}
which gives $p(\kappa = 0 | d) = 0.986$ for the Astronomer's prior, starting from a prior probability $p(\kappa = 0) = 3/9 \approx 0.33$ (as there are 3 flat models among the 9 models we consider here). For the more conservative Curvature scale prior, we obtain a posterior probability for flatness of only $p(\kappa = 0 | d) = 0.462$. The origin of this difference can once more be traced back to the less pronounced Occam's razor effect against non-flat models for the Curvature scale prior. 

\begin{table*}
\begin{center}\begin{tabular}{ | l | l l l l l l l l | l |}
\hline
Prior & CWZ & CW & C$\Lambda$ & OWZ & OW & O$\Lambda$ & FWZ & FW & $p(\kappa = 0 | d)$ \\
\hline 
Astronomer's & $8.28 \pm 0.09$ & $6.60 \pm 0.09$ & $4.61\pm 0.09$ & $7.41\pm0.06$ & $7.07\pm 0.05$ & $5.63\pm 0.03$ & $2.67\pm 0.03$ & $2.13\pm 0.03$ & $0.986 \pm 0.005$\\
Curvature & $3.28\pm 0.1$ & $2.79\pm0.1$ & $0.73\pm0.04$ & $2.82\pm0.05$ & $2.31\pm0.04$ & $0.44\pm0.04$ & $2.67\pm 0.03$ & $2.13 \pm 0.03$ & $0.462 \pm 0.006$\\
\hline
\end{tabular}
\end{center}
\caption{Log of the Bayes factors ($\ln B$) between the flat $\Lambda$CDM model and alternative models with curvature and/or an evolving dark energy equation of state (indicated by the label in the top line), for both our choices of priors for the curvature sector. The last column gives the posterior probability of the Universe being flat. }
\label{tab:model_selection}
\end{table*}

\begin{figure}
\begin{center}
\includegraphics[width=0.49\linewidth]{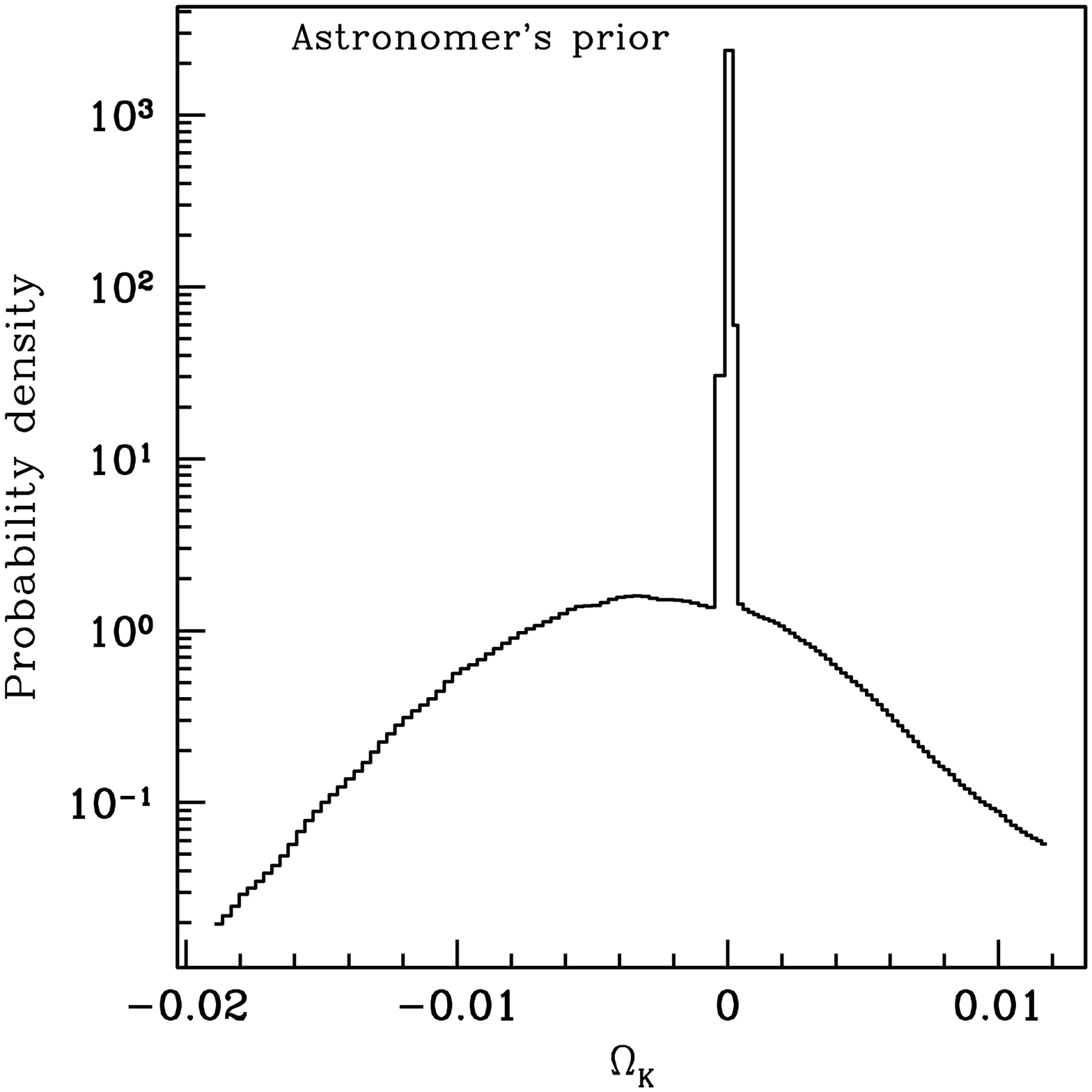}
\includegraphics[width=0.49\linewidth]{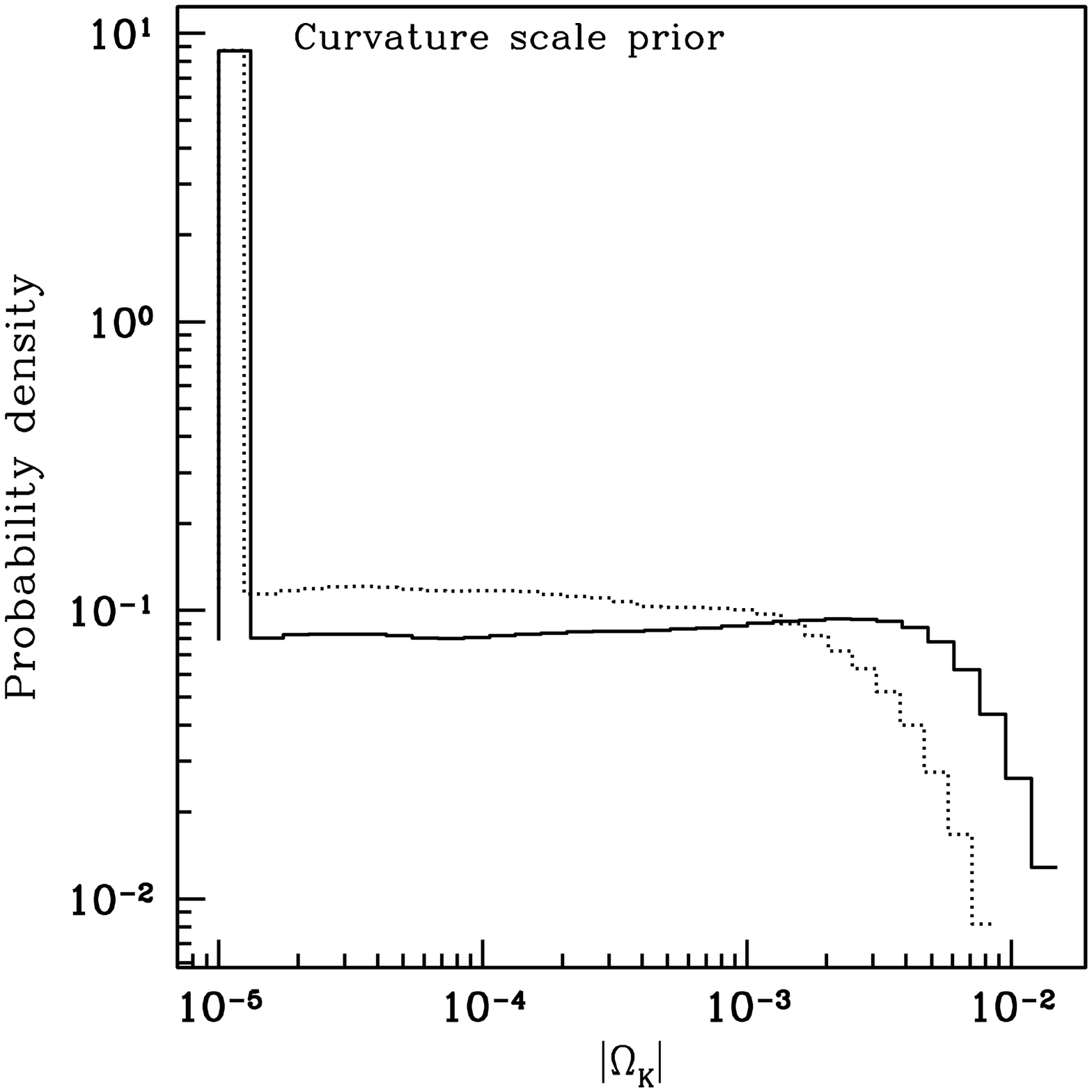}
\caption{Model-averaged posterior probability distribution for the curvature parameter, including all 9 models considered in the analysis, assuming the Astronomers' prior (left panel) and the Curvature scale prior (right panel) for the curvature parameter. In the right panel, the solid line applies to closed Universes ($\Omk < 0$), while the dotted line to open Universes ($\Omk > 0$). The peaks represent the Dirac delta function encompassing the probability mass associated with flat models.}
\label{fig:BMAcurvature}
\end{center}
\end{figure}

Using the values of the Bayes factors given in Table~\ref{tab:model_selection}, we derive in Table~\ref{tab:constraints} model-averaged constraints on the quantities of interest ($\Omk, N_U, R_c$).  The table also gives the non-model averaged constraints for the most conservative model (with a possibly time-evolving dark energy equation of state) for comparison. The model-averaged posterior for the curvature parameter is also plotted in Fig.~\ref{fig:BMAcurvature} for both choices of priors, where the spike in both panels represents the posterior probability mass concentrated in the flat models ($\sim 99\%$ for the Astronomer's prior and $\sim 46\%$ for the Curvature scale prior). Because the posterior probability of flat models gets concentrated in a delta-function at $\Omk = 0$ (for the Astronomer's prior) or $|\Omk| = 10^{-5}$ (for the Curvature scale prior), the model-averaged constraints on $\Omk$ can produce {\em tighter} intervals than the usual results obtained under the assumption that $\Omk\neq 0$. This ``concentration of probability'' onto the simpler model is a consequence of the Occam's razor effect implicit in Bayesian model selection. A similar effect has been observed in model-averaged constraints on the dark energy equation of state~\citep{Liddle:2006kn}. Because more than 95\% of the posterior probability is concentrated in flat models under the Astronomer's prior, 95\% limits on the curvature parameter are not defined for this prior. At 99\%, we find that $|\Omk| \leq 2\times10^{-4}$, while the number of Hubble spheres is greater than 398 and the radius of curvature larger than 68 Gpc (all at 99\%). The Curvarture  scale prior is more conservative, in that it penalizes less strongly non-flat models. As a consequence, we find that $-0.9\times 10^{-2} \leq \Omk \leq 1.0 \times 1.0^{-2}$ (99\% model-averaged region), while the number of Hubble spheres is $N_U \gsim 251$ and the curvature radius $R_c \gsim 42$ Gpc. Even under this more conservative prior, the model-averaged constraint on curvature is a factor $\sim 2$ better than the usual, non model-averaged results, which (from a similar collection of data) is  approximately $|\Omk| \leq 1.7\times10^{-2}$ at 99\%~\citep{Komatsu:2010fb}. The model-averaged constraint on the number of Hubble spheres is a factor $\sim 40$ stronger than the non model-averaged one: in the latter case, and using the Curvature scale prior, \cite{Vardanyan:2009ft} found $N_U \gsim 6.2$ (at 99\%).

\begin{table}
\begin{center}\begin{tabular}{| l | ll  | ll | }
\hline
Quantity & \multicolumn{2}{c|}{Astronomers' prior} &  \multicolumn{2}{c|}{Curvature scale prior} \\
		   & 95\%  & 99\%  &  95\%  & 99\% \\ 
\hline
\multicolumn{5}{|c|}{Model-averaged constraints}  \\ 
\hline
$\Omk \times 10^{-3}$ & N/A & [-0.2,0.2] & [-4.4, 2.0] & [-8.9, 9.9]\\ 
$\curv$ (Gpc) & N/A & $>68$ & $>88$ & $>42$\\
$N_U$ & N/A & $>398$ & $>1000$ & $>251$\\
\hline   
\multicolumn{5}{|c|}{Non model-averaged constraints (WZ model)}  \\
\hline

$\Omk \times 10^{-3}$ & [-14,9] & [-19,17] & [-9, 7] & [-14, 13]\\
$\curv$ (Gpc) & $>34$ & $>32$ & $>30$ & $>28$\\
$N_U$ & $>33$ & $>29$ & $>38$ & $>32$\\
\hline
\end{tabular} \caption{Top section: model-averaged parameter constraints for the curvature parameter, the curvature scale and the number of Hubble spheres in the Universe. For the Astronomer's prior, $\sim 97\%$ of the posterior probability is concentrated in the flat models, and therefore the 95\% limit for the model-averaged parameters is not defined. Bottom section: non model-averaged constraints for the most conservative case (evolving dark energy model), for comparison. \label{tab:constraints} }
\end{center}
\end{table}

\section{Conclusions}
\label{sec:conclusions}

We have applied the formalism of Bayesian model averaging to the problem of constraining the curvature and size of the Universe. By employing the Savage-Dickey density ratio, we have obtained model-averaged constraints at almost no additional computational effort than what is needed for parameter estimation. We have demonstrated how model-averaged constraints on the curvature and minimum size of the Universe can be considerably tigther than non model-averaged ones. This is a consequence of the fact that flat models are preferred by Bayesian model selection, although the strength of such preference is fairly strongly dependent on the choice of prior for the curvature parameter.

We have considered two classes of priors that are based on physical and theoretical considerations. We found that even the most conservative prior choice gives model-averaged constraints on curvature that are a factor of $\sim 2$ better than non model-averaged intervals. A more aggressive prior choice (the Astronomer's prior) leads to an improvement in the constraints on $\Omk$ by a factor $\sim 100$, giving $|\Omk| \leq 2\times10^{-4}$ at 99\%. The minimum size of the Universe is robustly constrained to encompass $N_U \gsim 251$ Hubble spheres, an improvement of a factor $\sim 40$ on previous constraints. Finally, the radius of curvature of spatial section is found to be $R_c \gsim 42$ Gpc.

Bayesian model averaging gives the most general parameter constraints, which fully account for the uncertainty in the selection of the correct underlying cosmological model. It remains imperative (like in any good Bayesian analysis) to study the dependency of the results of the chosen priors, which are more important in Bayesian model selection (and model averaging) than they are in the usual parameter inference framework. We believe that the formalism presented here can be employed successfully in a large variety of cosmological problems. 

\bigskip

\textit{Acknowledgements.} The authors would like to thank Andrew Jaffe and Andrew Liddle for useful discussions. R.T. would like to thank the Institut d'Astrophysique de Paris for hospitality.
M.V. is supported by the Raffy Manoukian Scholarship and partially supported by the Philip Wetton Scholarship at Christ Church, Oxford. 
 We acknowledge the use of the Legacy Archive
for Microwave Background Data Analysis (LAMBDA). Support for
LAMBDA is provided by the NASA Office of Space Science.

%\bibliographystyle{mn2e}
%\bibliography{References}

\begin{thebibliography}{}
\bibitem[\protect\citeauthoryear{Clarkson et~al.,}{Clarkson
  et~al.}{2007}]{Clarkson:2007bc}
  C.~Clarkson, M.~Cortes, B.~A.~Bassett,
  %``Dynamical Dark Energy or Simply Cosmic Curvature?,''
  JCAP {\bf 0708}, 011 (2007).

\bibitem[\protect\citeauthoryear{Dunkley et~al.,}{Dunkley
  et~al.}{2009}]{Dunkley:2008ie}
Dunkley J.,  et~al., 2009, Astrophys. J. Suppl., 180, 306

\bibitem[\protect\citeauthoryear{Freedman et~al.,}{Freedman
  et~al.}{2001}]{Freedman:2000cf}
Freedman W.~L.,  et~al., 2001, Astrophys. J., 553, 47

\bibitem[\protect\citeauthoryear{Hicken et~al.,}{Hicken
  et~al.}{2009}]{Hicken:2009dk}
Hicken M.,  et~al., 2009, Astrophys. J., 700, 1097

\bibitem[\protect\citeauthoryear{Komatsu et~al.,}{Komatsu
  et~al.}{2009}]{Komatsu:2008hk}
Komatsu E.,  et~al., 2009, Astrophys. J. Suppl., 180, 330

\bibitem[\protect\citeauthoryear{Komatsu et~al.,}{Komatsu
  et~al.}{2011}]{Komatsu:2010fb}
Komatsu E.,  et~al., 2011, Astrophys. J. Supp., 192, 18

\bibitem[\protect\citeauthoryear{Kowalski et~al.,}{Kowalski
  et~al.}{2008}]{Kowalski:2008ez}
Kowalski M.,  et~al., 2008, Astrophys. J., 686, 749

\bibitem[\protect\citeauthoryear{Kunz, Trotta \& Parkinson}{Kunz
  et~al.}{2006}]{Kunz:2006mc}
Kunz M.,  Trotta R.,    Parkinson D.,  2006, Phys. Rev., D74, 023503

\bibitem[\protect\citeauthoryear{Liddle, Mukherjee, Parkinson \& Wang}{Liddle
  et~al.}{2006}]{Liddle:2006kn}
Liddle A.~R.,  Mukherjee P.,  Parkinson D.,    Wang Y.,  2006, Phys. Rev., D74,
  123506

\bibitem[\protect\citeauthoryear{March, Starkman, Trotta \& Vaudrevange}{March
  et~al.}{2010}]{March:2010ex}
March M.~C.,  Starkman G.~D.,  Trotta R.,    Vaudrevange P.~M.,  2010

\bibitem[\protect\citeauthoryear{Marshall, Hobson \& Slosar}{Marshall
  et~al.}{2003}]{Marshall:2003ez}
Marshall P.~J.,  Hobson M.~P.,    Slosar A.,  2003, Mon. Not. Roy. Astron.
  Soc., 346, 489

\bibitem[\protect\citeauthoryear{Parkinson \& Liddle}{Parkinson \&
  Liddle}{2010}]{Parkinson:2010zr}
Parkinson D.,  Liddle A.~R.,  2010, Phys. Rev., D82, 103533

\bibitem[\protect\citeauthoryear{Percival et~al.,}{Percival
  et~al.}{2007}]{Percival:2007yw}
Percival W.~J.,  et~al., 2007, Mon. Not. Roy. Astron. Soc., 381, 1053

\bibitem[\protect\citeauthoryear{Reid et~al.,}{Reid
  et~al.}{2010}]{Percival:2009xn}
Reid B.~A.,  et~al., 2010, Mon. Not. Roy. Astron. Soc., 401, 2148

\bibitem[\protect\citeauthoryear{Scott \& Zibin}{Scott \&
  Zibin}{2006}]{Scott:2006kga}
Scott D.,  Zibin J.~P.,  2006, Int. J. Mod. Phys., D15, 2229

\bibitem[\protect\citeauthoryear{Starkman, Trotta \& Vaudrevange}{Starkman
  et~al.}{2008}]{doubt}
Starkman G.,  Trotta R.,    Vaudrevange P.~M.,  2008, arXiv:0811.2415

\bibitem[\protect\citeauthoryear{Trotta}{Trotta}{2007}]{Trotta:2005ar}
Trotta R.,  2007, Mon. Not. Roy. Astron. Soc., 378, 72

\bibitem[\protect\citeauthoryear{Trotta}{Trotta}{2008}]{Trotta:Bayes}
Trotta R.,  2008, Contemporary Physics, 49, 71

\bibitem[\protect\citeauthoryear{Vardanyan, Trotta \& Silk}{Vardanyan
  et~al.}{2009}]{Vardanyan:2009ft}
Vardanyan M.,  Trotta R.,    Silk J.,  2009, Mon. Not. Roy. Astron. Soc., 397,
  431

\bibitem[\protect\citeauthoryear{Verdinelli \& Wasserman}{Verdinelli \&
  Wasserman}{1995}]{Verdinelli:1995}
Verdinelli I.,  Wasserman L.,  1995, J. Amer. Stat. Assoc., 90, 614

\end{thebibliography}

\end{document}